\documentclass[manuscript]{aastex6}

\usepackage{mathtools}
\usepackage{epsfig}
\usepackage{comment}
\fullcollaborationName{The Friends of Fermi Asian Network}

\begin{document}


\title{Gamma-ray Emission of the Kes 73/1E 1841-045 Region Observed with the \emph{Fermi} Large Area Telescope}

%% Use \author, \affil, plus the \and command to format author and affiliation 
%% information.  If done correctly the peer review system will be able to
%% automatically put the author and affiliation information from the manuscript
%% and save the corresponding author the trouble of entering it by hand.
%%
%% The \affil should be used to document primary affiliations and the
%% \altaffil should be used for secondary affiliations, titles, or email.

%% Authors with the same affiliation can be grouped in a single
\author{Paul K. H. Yeung\altaffilmark{1}} 

\author{Albert K. H. Kong\altaffilmark{1}}

\author{P. H. Thomas Tam\altaffilmark{2}}

\author{C. Y. Hui\altaffilmark{3}}

\author{Jumpei Takata\altaffilmark{4}}

\and

\author{K. S. Cheng\altaffilmark{5}}

\altaffiltext{1}{Institute of Astronomy and Department of Physics, National Tsing Hua University, Hsinchu, Taiwan; \email{paul2012@connect.hku.hk}, \email{akong@phys.nthu.edu.tw}}
\altaffiltext{2}{Institute of Astronomy and Space Science, Sun Yat-Sen University, Guangzhou 510275, China; \email{tanbxuan@mail.sysu.edu.cn}}
\altaffiltext{3}{Department of Astronomy and Space Science, Chungnam National University, Daejeon 305-764, Republic of Korea}
\altaffiltext{4}{School of physics, Huazhong University of Science and Technology, Wuhan 430074, China}
\altaffiltext{5}{Department of Physics, University of Hong Kong, Pokfulam Road, Hong Kong}

\begin{abstract}

The supernova remnant Kes 73 and/or the magnetar 1E 1841-045 at its center can deposit  a large amount of energy to the surroundings and is potentially responsible for particle acceleration. Using the data taken with the \emph{Fermi} Large Area Telescope (LAT), we confirmed the presence of an extended source whose  centroid position is highly consistent with this magnetar/supernova-remnant pair. Its  emission is  intense from 100 MeV to $>$100 GeV. Its LAT spectrum can be decoupled into two components which are respectively governed by two different mechanisms. According to the young age of this system, the magnetar is seemingly a necessary and sufficient source for the downward-curved spectrum below 10 GeV, as the observed $<$10 GeV flux is too high for the supernova remnant to account for. On the other hand, the supernova remnant  is reasonably responsible for the hard  spectrum above 10 GeV. Further studies of this region in the TeV regime is required, so that we can perform physically meaningful comparisons of the $>$10 GeV spectrum and the TeV spectrum. %An apparent  flux drop at the GeV band during a 720-day interval (2011 January 21$-$2013 January 10) is also reported. In another region, the supernova remnant Kes 79 is likely to interact hadronically with its surrounding H II region, which is positionally consistent with what we observe with  LAT: another point-like source with a steady flux and a broken-power-law spectrum. There are three pulsars, including the magnetar 3XMM J185246.6+003317, inside this supernova remnant or in its neighbourhood. We discuss possible origins of these two sources by considering both supernova remnants' and pulsars' contributions.

\end{abstract}

\keywords{pulsars: individual (1E 1841-045) $-$ stars: magnetars  $-$ ISM: individual objects (SNR  Kes 73, HESS J1841-055) $-$ ISM: cosmic rays $-$ gamma rays: general}



\section{Introduction} \label{sec:intro}

Magnetars are neutron stars with typically longer periods of spinning ($\sim$2$-$12 s) and  stronger surface magnetic fields ($>$$6\times10^{12}$ G)~\footnote{McGill magnetar catalog: \url{http://www.physics.mcgill.ca/\~pulsar/magnetar/main.html}} \citep[][]{Olausen2014},  compared to normal neutron stars. Their manifestations include soft gamma-ray repeaters (SGRs) and anomalous X-ray pulsars (AXPs), which are mostly located at low Galactic latitudes \citep[][]{Olausen2014}. Magnetars emit mainly by  releasing magnetic energy rather than rotational energy \citep[][]{Duncan1992}. Due to unusually high surface magnetic fields, frequent starquakes occur on magnetars \citep[][]{Duncan1998}, leading to recurrent X-ray outbursts of magnetars \citep[cf.][and references therein~\footnote{\label{GCN}GCN Circulars: \url{http://gcn.gsfc.nasa.gov/gcn3\_archive.html}}]{Collazzi2015}. Whenever a magnetar outburst takes place, a substantial amount of energy spills, following the ejecta, to the surroundings and can affect the emission in that region at energies above 1~keV.  Whether the gigaelectronvolt (GeV)-emissions from regions of magnetars react correspondingly to magnetar outbursts or not deserves to be a focus of our attention.

Pulsed GeV radiation from the magnetospheres of magnetars have been predicted by theoretical models \citep[e.g.,][]{Cheng2001, Beloborodov2007, Takata2013}. Nonetheless, it is arguable that the GeV pulsations of magnetars are likely to be buried  due to their low flux levels \citep[cf.][]{Li2016}. The difficulties in pulsation search of magnetars in $\gamma$-ray are further increased by the  complicated $\gamma$-ray backgrounds in magnetars' neighbourhoods, because the Galactic diffuse emission is  strong and the distributions of celestial objects are  crowded at Galactic latitudes within $\pm$1$^\circ$. Celestial objects in magnetars' neighbourhoods include, but are not limited to, magnetars' derivative products.  While analyses of \emph{Fermi} Large Area Telescope (LAT) data did not give convincing evidence for $\gamma$-rays from any magnetar \citep[cf.][]{Abdo2010d, Auchettl2014, Yeung2016, Li2016}, it is  interesting to observe $\gamma$-rays from some derivative products of magnetars,  as some of them are also magnetically powered and enable us to study physics of ultra-high magnetic field.

Supernova remnants (SNRs) are common derivative products of magnetars as well as other categories of neutron stars, and form through stellar collapse and subsequent supernova explosion. Thanks to LAT observations, plenty of GeV-bright SNR$-$molecular cloud (MC) association systems have been significantly detected  \citep[e.g.,][]{Abdo2009, Abdo2010a, Abdo2010b, Abdo2010c, Castro2013, Xing2014, Liu2015, Araya2015} and discovered to have broken-power-law or log-parabola spectra \citep[][]{Acero2015b}, suggesting an escape of over-energetic particles from the acceleration sites. Pulsar wind nebulae (PWNe) are another type of pulsars' derivative products, which form through pulsars' wind materials interacting with and shocked by ambient medium. They have been studied from the radio band \citep[synchrotron radiation; e.g.,][]{Ma2016} to the $\gamma$-ray band. At least 27 pulsars have their PWNe identified as teraelectronvolt (TeV) sources \citep[cf.][]{Halpern2010b} emitting through inverse-Compton (IC) scattering \citep[reviewed by][]{Gaensler2006}. Until now, Swift J1834.9-0846 is the only magnetar confirmed to produce an X-ray PWN \citep[][]{Younes2016},  and CXOU J171405.7-381031  is the only magnetar argued as producing a TeV PWN \citep[][]{Halpern2010b}. 

1E 1841-045 is an AXP  discovered to be the  central source of SNR Kes 73 \citep[][]{Vasisht1997}. The  Kes 73/1E 1841-045 system has a distance of $8.5^{+1.3}_{-1.0}$  kpc from us and an age of  750-2100 yr \citep[][]{Tian2008, Kumar2014}. This system is potentially responsible for particle acceleration, as it shows intense TeV emission at the northern edge of  HESS J1841-055 \citep[][]{Aharonian2008, Bochow2011}. This arouses our motivation to search for the GeV counterpart of the magnetar/SNR pair. 1E 1841-045  has a spin-down power of  $\sim$$1\times10^{33}$ erg s$^{-1}$ and a surface magnetic field of $\sim$$7\times10^{14}$ G \citep[][]{Dib2014}. 1E 1841-045 experienced a series of 10 X-ray outbursts in 2011 February$-$July \citep[][]{Lin2011}, and it is interesting to see how the GeV flux varies shortly after these outbursts. Tight interaction of Kes 73 with MCs is supported by detection of  OH (1720 MHz) maser emission \citep[][]{Green1997}, discovery of the methanol source  IRAS 18379-0500 which is associated with an  H II region \citep[][]{Helfand1992, Slysh1999}, and revelation of broad-molecular-line (BML) regions by H I and CO observations \citep[][]{Tian2008, Kilpatrick2016}. Nevertheless, the total solid angle subtended at Kes 73 by its adjacent compact H II regions is only $\sim$5\% of an entire sphere \citep[cf. Figure 1(a) of][]{Tian2008} and, to some degree, limits the $\gamma$-ray contribution from Kes 73.

In this paper, we explore the hundred-megaelectronvolt-to-GeV emission in the field of the Kes 73/1E 1841-045 system by using $\sim$7.5 years of \emph{Fermi} LAT data with the latest instrumental responses and background models. %We do \emph{not} expect Kes 79, 3XMM J185246.6+003317 and two other pulsars to be well resolved by LAT either, assuming they are both strong GeV sources. 
Then, we  provide some insight into the possible origin(s) of the $\gamma$-rays, in terms of the spectral shapes, energy budgets  and long-term temporal behaviors.  Results of a nearby region consisting of 3XMM J185246.6+003317 will be published in a follow-up paper.

\section{Observation \& Data Reduction} \label{sec:data}

We performed a series of binned maximum-likelihood analyses for a 20$^\circ$$\times$20$^\circ$ region-of-interest (ROI) centered at RA=$18^{h}47^{m}04.963^{s}$, Dec=$-02^\circ11'37.66"$ (J2000), which is approximately the midpoint between  1E 1841-045 and 3XMM J185246.6+003317. We  used the data obtained by LAT between 2008 August 4 and 2016 February 1. The data were reduced and analyzed with the aid of \emph{Fermi} Science Tools v10r0p5 package. In view of the complicated environment of the Galactic plane regions, we adopted the events classified as Pass8 ``Clean" class for the analysis so as to better suppress the background. The corresponding instrument response function (IRF) ``P8R2$_-$CLEAN$_-$V6" is used throughout the investigation. We further filtered the data by accepting only the good time intervals where the ROI was observed at a zenith angle less than 90$^\circ$ so as to reduce the contamination from the albedo of Earth.

For subtracting the background contribution, we  included the Galactic diffuse background (gll$_-$iem$_-$v06.fits), the isotropic background (iso$_-$P8R2$_-$CLEAN$_-$V6$_-$PSF3$_-$v06.txt for ``PSF3" data, iso$_-$P8R2$_-$CLEAN$_-$V6$_-$FRONT$_-$v06.txt for ``FRONT" data or iso$_-$P8R2$_-$CLEAN$_-$V6$_-$v06.txt for a full set of data) as well as all other  point sources cataloged in the most updated Fermi/LAT catalog \citep[3FGL;][]{Acero2015a} within 25$^\circ$ from the ROI center in the source model.  In \S\ref{sec:spatial}, we report discoveries of a few  sources within 5$^\circ$ from the ROI center, which were also included in the source model. We  set free the spectral parameters of the  sources within 8$^\circ$ from the ROI center in the analysis. For the   sources beyond 8$^\circ$ from the ROI center, their spectral parameters were fixed at the catalog values. 

In spectral and temporal analysis, we required each energy-bin and time-segment to attain a signal-to-noise ratio $\gtrsim$$3.0\sigma$  (equivalently, a TS value $\gtrsim$9 and a chance probability $\lesssim$0.3\%) for a robust result. For each energy-bin or time-segment \emph{dissatisfying} this requirement, we placed  a 3.0$\sigma$ upper limit on its flux.

\section{Data Analysis} \label{sec:results}

\subsection{Spatial Analysis} \label{sec:spatial}

We investigate the morphology of the Kes 73/1E 1841-045 region in high-energy (HE; $>$1 GeV) and very-high-energy (VHE; $>$10 GeV) regimes respectively. We notice that the PSF at 10 GeV for a full set of data is smaller than that at 1 GeV for ``PSF3" data (cf. SLAC~\footnote{\label{slac}Fermi LAT Performance: \url{http://www.slac.stanford.edu/exp/glast/groups/canda/lat_Performance.htm}}). Therefore, in order to achieve a compromise between good spatial resolution and adequate photon statistics, we independently analyze two sets of data for morphological studies: (1) ``PSF3" data in 1$-$50 GeV, (2) and a full set of data in 10$-$200 GeV.

\begin{figure}
	\centering
	\includegraphics[width=0.49\linewidth]{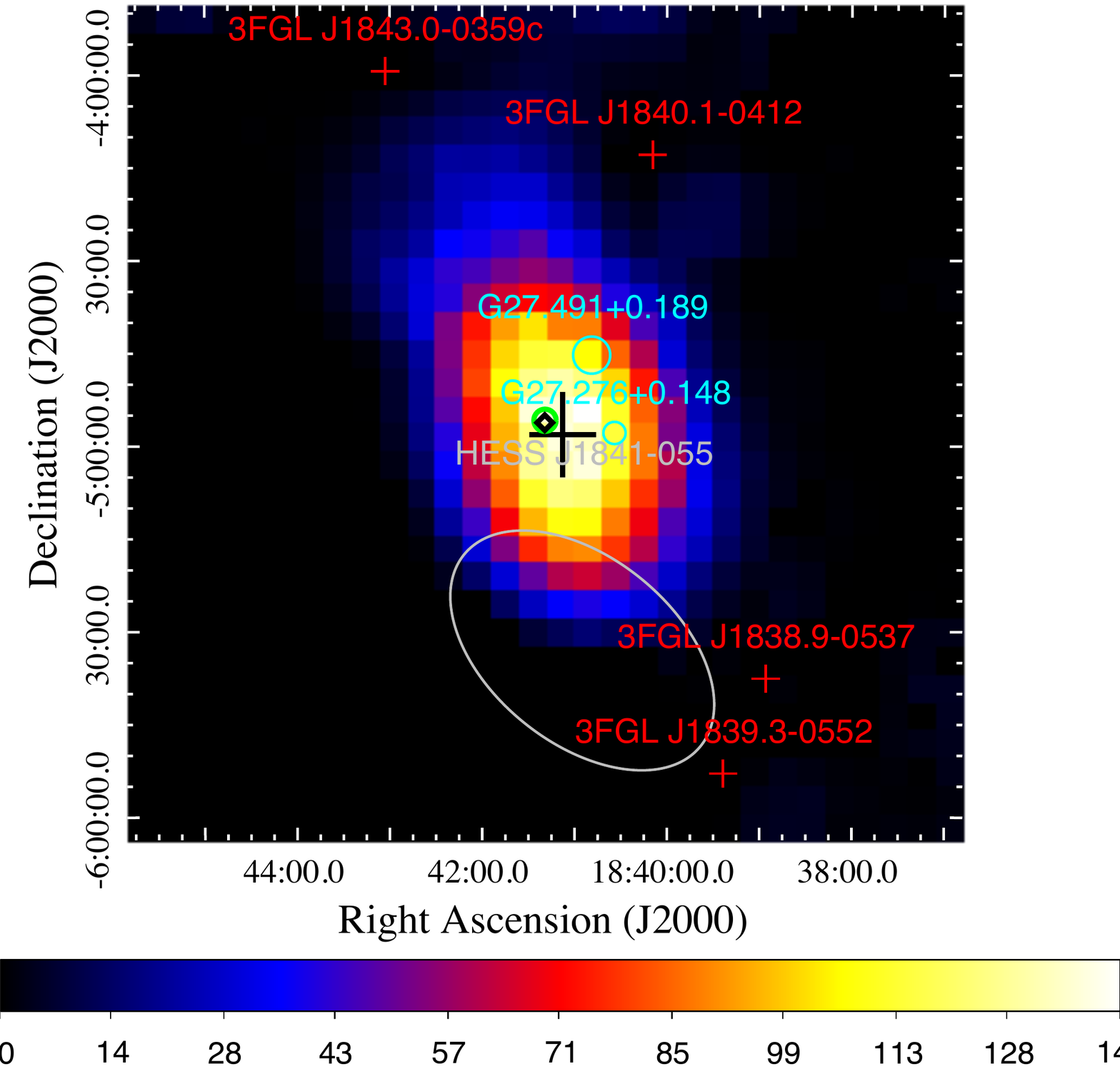}
	\includegraphics[width=0.49\linewidth]{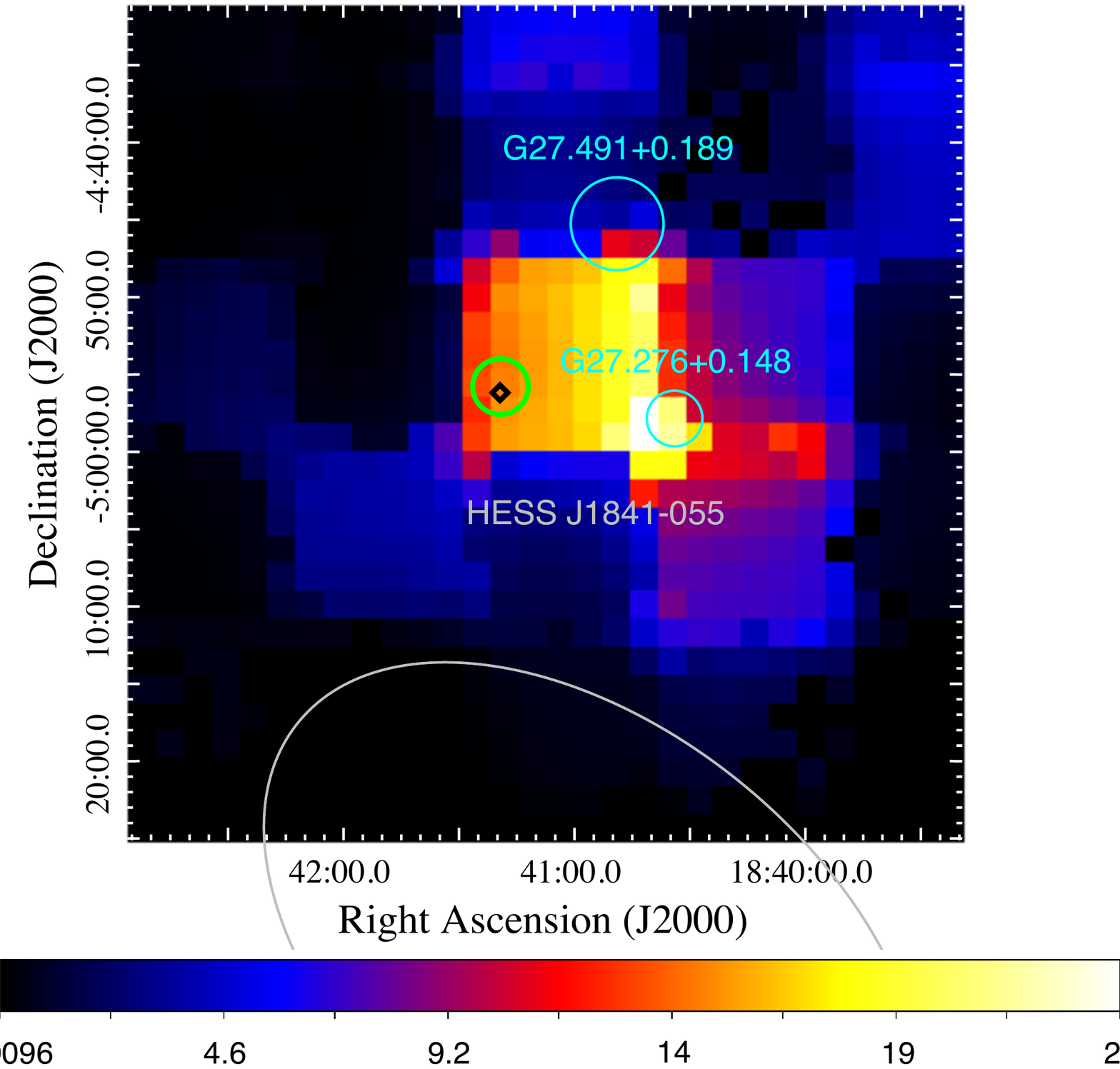}
	\caption{TS maps of the field around 1E 1841-045 in 1$-$50 GeV (for ``PSF3" data; left) and 10$-$200 GeV (for a full set of data; right), respectively, where all neighboring 3FGL catalog sources are subtracted. The green thick circles indicate the position and dimension of Kes 73, which are taken from \citet{Acero2015b}. The black diamonds indicate the position of 1E 1841-045, which is taken from \citet{Wachter2004}. The position and dimension of the two H II clouds G27.276+0.148 and G27.491+0.189,  indicated as cyan circles, are taken from \citet{Tian2008}. The position and dimension of HESS J1841-055,  indicated as a gray ellipse, are taken from \citet{Aharonian2008}.  The positions of nearby 3FGL sources are marked by red crosses. On the left map, the centroid  is indicated as a black thick cross, whose size illustrates the 95\% error region.}
	\label{1E_tsmap}
\end{figure}

The test-statistic (TS) maps of the field around  1E 1841-045  are shown in Figure~\ref{1E_tsmap}, where all 3FGL catalog sources including HESS J1841-055 are subtracted.   The peak detection significance is $\sim$$11.9\sigma$ in 1$-$50 GeV and $\sim$$4.8\sigma$ in 10$-$200 GeV.  On the TS map in 1$-$50 GeV, we determined the contour where the TS value is lower than the maximum by 9.488. Since this contour is very ellipse-like, we fit an ellipse to it. The center of this ellipse is determined to be the best-fit centroid, while its dimension is determined to be the  95\% error region for 4 d.o.f.~\footnote{Ones can refer to the Chi-Square Distribution Table, which shows that $\chi_{.050}^{2}=9.488$ for 4 d.o.f.. There are 4 d.o.f., because of 4 variables: right ascension, declination, flux normalization and photon index.}. We found that the 95\% error region of the   1$-$50 GeV centroid covers the entire  Kes 73/1E 1841-045 system, but is mutually exclusive with its two adjacent  compact H II  regions,  G27.276+0.148 and G27.491+0.189. The best-fit 10$-$200 GeV centroid is offset from the magnetar/SNR pair by $\sim$$9'$, but its 95\% error region contains the pair. The spatial coincidence of the 1$-$50 GeV feature LAT observed and HESS J1841-055 is very marginal, but the  1$-$50 GeV centroid determined from LAT data still has  a detection significance of $\sim$$6.0\sigma$ in HESS \citep[cf.][]{Aharonian2008, Bochow2011}. 

%The 1$-$50 GeV TS map of the field around  3XMM J185246.6+003317 for ``PSF3" data are shown in Figure~\ref{3XMM_tsmap}, where all 3FGL catalog sources are subtracted. The morphology is point-like. The 95\% error region of the   1$-$50 GeV centroid excludes  3XMM J185246.6+003317,  PSR B1849+00 as well as the  Kes 79/CXOU J185238.6+004020 system, while well overlapping with their surrounding H II  region. The 10$-$200 GeV background-subtracted TS map of this field shows that the detection significance is $\lesssim$$3\sigma$ everywhere within the H II region. HESS observations of   Kes 79 yield a detection significance of only $\sim$$2.7\sigma$ in the TeV band \citep[][]{Bochow2011}.

For  examining whether the morphology of the emission from this region is extended, we followed a scheme adopted by  \citet{Yeung2016}. %We produced a 1$-$50 GeV count-map where all 3FGL catalog sources including HESS J1841-055 are subtracted, and then computed a brightness profile along the seemingly extended orientation. We also simulated the expected point-like source with the same spectrum as this region. The result is shown in Figure~\ref{bgsub}. %Considering some slight mis-match between the simulation-based IRF and the actual in-orbit performance, we consider a source to be extended only if its FWHM exceeds that of the simulated point source by $\ge$$3.0\sigma$.
%The brightness profile of the 1E 1841-045 region is well described by a single Gaussian, with a FWHM of $1.04^\circ \pm 0.25^\circ$ ($\chi^2=1.96$ for 9 d.o.f.),  exceeding that of the simulated point source, $0.47^\circ$, by  $\sim$$2.3\sigma$. {\bf This provides a marginal evidence for the emission from this region to be extended.} %Similarly, a single Gaussian provides a good fit to the brightness profile of the  3XMM J185246.6+003317 region, where the  FWHM is $1.22^\circ \pm 0.30^\circ$ ($\chi^2=2.11$ for 8 d.o.f.),  exceeding that of the simulated point source, $0.43^\circ$, by only $<$$2.7\sigma$.
%Due to the large uncertainties resulted from the noisy background,  there is \emph{no} conclusive evidence for the emission from these two regions to be extended, and we modelled them as two additional point sources respectively, in subsequent analyses. 
We  performed a likelihood ratio test to quantify the significance of extension. We adopted the position of the  1$-$50 GeV centroid, (18:41:07.723, $-$04:58:07.44 (J2000)), for the  1E 1841-045 region (namely \emph{Fermi} J1841.1-0458). We assigned it a simple power law, and we attempted  uniform-disk morphologies of different radii as well as a point-source model on it. We took the ``FRONT" data in 1$-$400 GeV, which thoroughly covers both HE and VHE bands. The $2\Delta$$ln(likelihood)$ of different radii relative to the point-source model are tabulated in Table~\ref{Ext}. The most likely radius is determined to be $0^\circ.32^{+0^\circ.05}_{-0^\circ.01}$~\footnote{The $1\sigma$ uncertainties are determined at where the $2\Delta$$ln(likelihood)$ is lower than the maximum by 1 according to the Chi-Square Distribution.} and this morphology is preferred over a point-source model by $\sim$$7.8\sigma$. We therefore adopted this morphology for \emph{Fermi} J1841.1-0458 in subsequent analyses. %(18:52:09.007, +0:36:52.36 (J2000)) for the  3XMM J185246.6+003317 region (namely \emph{Fermi} J1852.2+0037). 

In the  3XMM J185246.6+003317 region, we identified another extended source (namely \emph{Fermi} J1852.2+0037), which was modelled together with \emph{Fermi} J1841.1-0458 and will be reported in our  follow-up paper. Within 5$^\circ$ from the ROI center, we  discovered two more  sources at   (281.79408$^\circ$, $-$2.3781283$^\circ$)$_{J2000}$ (TS$>$300) and (281.41503$^\circ$, $-$3.0315034$^\circ$)$_{J2000}$  (TS$>$50). In the same chain of analyses, we also modelled them as additional point sources, merely for the sake of better subtracting the $\gamma$-ray backgrounds.

\subsection{Spectral Analysis} \label{sec:spectral}

To construct the binned spectrum of \emph{Fermi} J1841.1-0458, we performed an independent fitting of each spectral bin.  For each spectral bin, we assigned a PL model to \emph{Fermi} J1841.1-0458. Considering that we include photons with energies below 1 GeV, and that we are investigating crowded regions in the Galactic plane, we find it \emph{inappropriate} to adopt a full set of data whose large PSF (e.g., a 68\% containment radius of $\sim$3.0$^\circ$ at 0.2 GeV; cf. SLAC~$^{\ref{slac}}$) leads to severe source confusion. Meanwhile, adopting only ``PSF3" data is also \emph{discouraged} in spectral fittings because of large systematic uncertainties induced by severe energy dispersion. For ``FRONT" data in 0.2$-$1 GeV, the FWHM of its PSF is $<$75\% of that for a full set of data and its energy dispersion is greater than that for a full set of data by only $\lesssim$1\% of the photon energy (cf. SLAC~$^{\ref{slac}}$). Therefore, we adopted only ``FRONT" data in order to achieve a compromise between a small PSF and lessened energy dispersion.

%\subsubsection{0.2$-$10 GeV spectra} \label{ref:0.2GeV}

\begin{figure}
	\centering
	\includegraphics[width=0.98\linewidth]{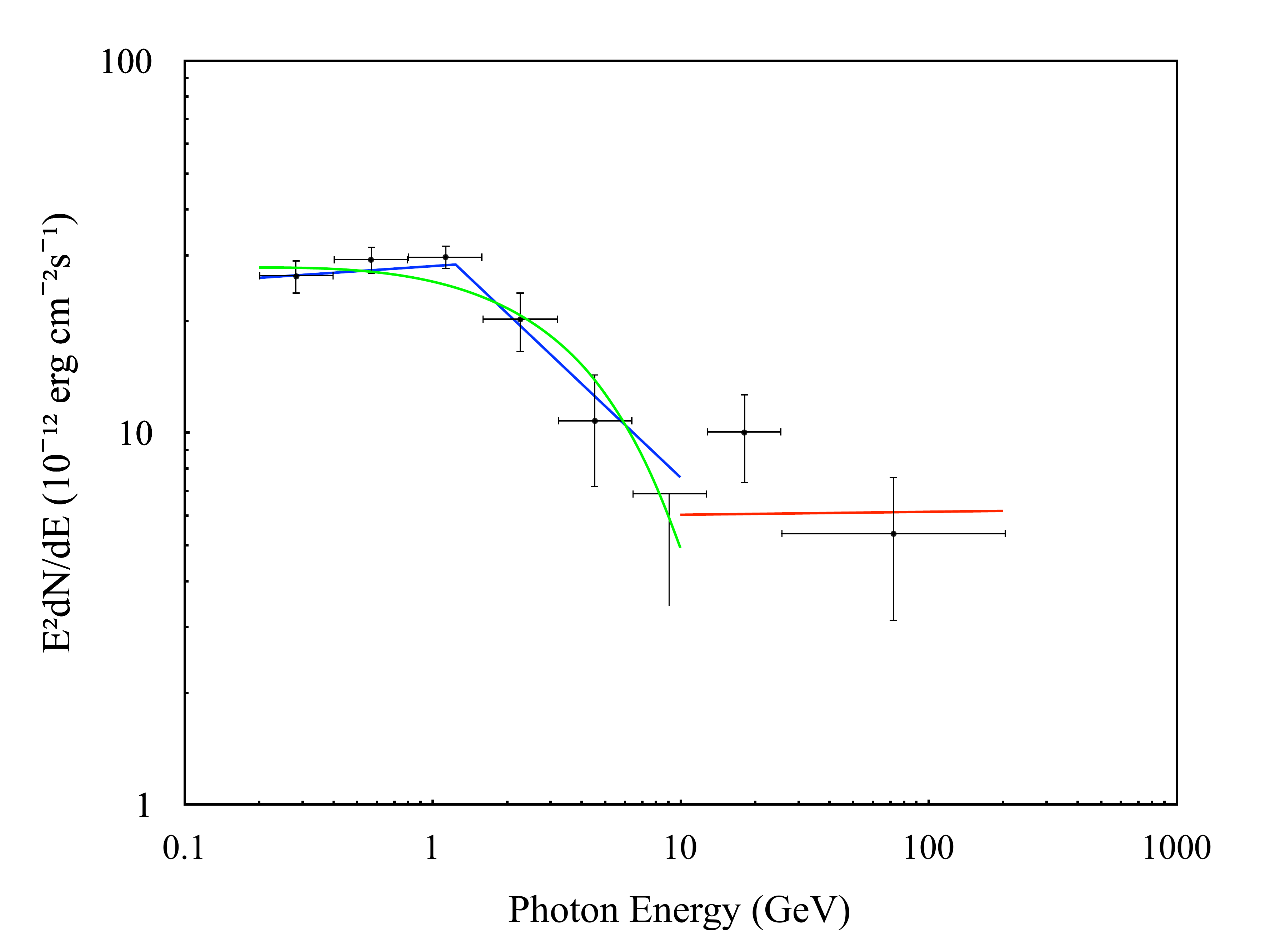}
	\caption{SED of \emph{Fermi} J1841.1-0458.  The upper limit is at  the $3.0\sigma$ confidence level. The green and blue  curves respectively  illustrate the best-fit PLE and BKPL models in 0.2$-$10 GeV, and the red line illustrates the best-fit  PL model in 10$-$200 GeV.  }
	\label{1E_SED}
\end{figure}

The spectral energy distribution (SED)  is shown in Figure~\ref{1E_SED}.  Noticeably, there appears to be a spectral disconnection at energies around 10 GeV, where the flux is  lower. This  suggests that the entire LAT spectrum may consist of two components, which are decoupled at energies around 10 GeV. With regards to this, we looked into the spectral shapes of two mutually exclusive energy bands respectively: 0.2$-$10 GeV and 10$-$200 GeV. We examined how well the spectrum in each band  can be described by, respectively, a simple power-law (PL)
\begin{center}
	$\frac{dN}{dE}=N_0(\frac{E}{E_0})^{-\Gamma}$ \ \ \ ,
\end{center}
an exponential cutoff power law  (PLE)
\begin{center}
	$\frac{dN}{dE}=N_0(\frac{E}{E_0})^{-\Gamma}\mbox{exp}(-\frac{E}{E_\mathrm{c}})$ \ \ \ ,
\end{center}
and a broken power law (BKPL)
\begin{center}
	$\frac{dN}{dE}=\begin{cases} N_0(\frac{E}{E_\mathrm{b}})^{-\Gamma_1} & \mbox{if } E<E_\mathrm{b} \\ N_0(\frac{E}{E_\mathrm{b}})^{-\Gamma_2} & \mbox{otherwise} \end{cases}$ \ \ \ .
\end{center}
The results of spectral fitting are tabulated in  Table~\ref{spectral1}.

For the  0.2$-$10 GeV spectrum of \emph{Fermi} J1841.1-0458, the likelihood ratio test indicates that both PLE and BKPL are preferred over PL by $>$$3.5\sigma$. A PLE model yields a photon index of $\Gamma=1.95 \pm 0.13$ and a cutoff energy of $E_\mathrm{c}=5092 \pm 2121$ MeV.  A BKPL model yields a photon index $\Gamma_{1}=1.95 \pm 0.08$ below the spectral break $E_\mathrm{b}=1241 \pm 249$ MeV and a photon index $\Gamma_{2}=2.63 \pm 0.11$ above the break.  Since the TS values yielded by PLE and BKPL models differ by $<$6, these two models are similarly preferable.

%The other target \emph{Fermi} J1852.2+0037 has a 0.2$-$10 GeV spectrum where  the likelihood ratio test indicates that both PLE and BKPL are preferred over PL by $\gtrsim$$3.0\sigma$. A PLE model yields a photon index of $\Gamma=2.44 \pm 0.19$ and a cutoff energy of $E_\mathrm{c}=2832 \pm 1656$ MeV.  A BKPL model yields a photon index $\Gamma_{1}=2.57 \pm 0.08$ below the spectral break, the spectral break $E_\mathrm{b}=986 \pm 115$ MeV and a photon index $\Gamma_{2}=3.20 \pm 0.16$ above the break. The spectrum  above $E_\mathrm{b}$ is steeper than that below $E_\mathrm{b}$ by $\sim$$3.6\sigma$. Since the TS values yielded by PLE and BKPL models differ by $<$1, these two models are equally preferable. 

%\subsubsection{10$-$200 GeV spectra} \label{ref:1GeV}

%$\S\ref{ref:0.2GeV}$ has revealed a break energy very close to 1.0 GeV for \emph{Fermi} J1841.1-0458. The previous fitting must be largely dominated by the photons at energies below the break. If we want to examine whether its spectrum has a second  break at an even higher energy, we inevitably need to exclude such dominating photons. This is why we repeated the spectral analysis with using 10$-$200 GeV data, in which almost all photons below the first turnover are excluded. We  examined how well the 10$-$200 GeV spectrum can be described by  PL and BKPL models respectively. The results are tabulated in  Table~\ref{spectral2} and  are overlaid in Figure~\ref{1E_SED}.

In the  10$-$200 GeV spectrum of \emph{Fermi} J1841.1-0458, a PL model yields a photon index $\Gamma=1.99 \pm 0.22$, which is consistent with a flat energy-spectrum within the tolerance of the $1\sigma$ uncertainty. The relatively low statistics do not allow distinguishing among the goodness of fit of PL, PLE and BKPL models. In order to demonstrate whether this  spectrum could be well connected to the 1$-$10 GeV spectrum by one single PL, we also performed a likelihood ratio test for the 1$-$200 GeV band. It turns out that a BKPL model with $E_\mathrm{b}$ fixed at 10 GeV ($\Gamma_{1}=2.65 \pm 0.10$ and $\Gamma_{2}=2.01 \pm 0.20$) is preferred over the best-fit PL ($\Gamma=2.53 \pm 0.09$) by $\sim$$2.8\sigma$ (i.e., the probability that the 1$-$200 GeV spectrum can be joined by a PL is $\sim$0.5\%). 

\subsection{Temporal Analysis} \label{sec:temporal}

\begin{figure}
	\centering
	\includegraphics[width=0.98\linewidth]{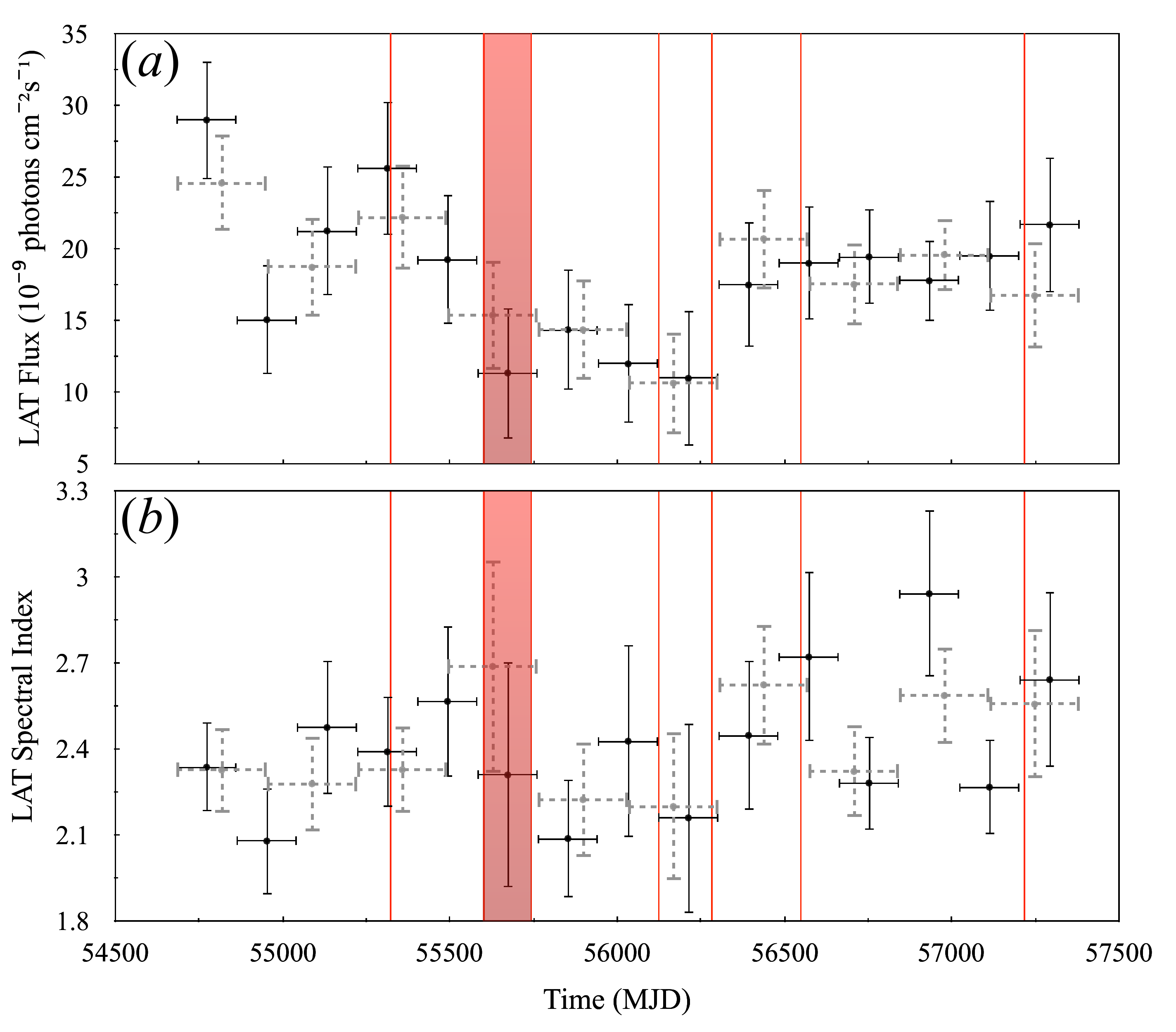}
	\caption{(a) 0.7$-$400 GeV light-curve of \emph{Fermi} J1841.1-0458. The size of each black solid bin is  180 days, and the size of each gray dashed bin is  270 days.  The red vertical lines indicate the dates of X-ray outbursts of  1E 1841-045, and the shaded time-range contains a series of 10 X-ray outbursts \citep[for detail, see][and references therein~$^{\ref{GCN}}$]{Lin2011, Collazzi2015}. (b) Temporal behavior of the 0.7$-$400 GeV spectral index of \emph{Fermi} J1841.1-0458 with  the same binning methods as those applied in (a).}
	\label{lc}
\end{figure}

In order to examine the long-term variability of \emph{Fermi} J1841.1-0458, we divided the first $\sim$7.4 years of \emph{Fermi} LAT observation into a number of 180-day segments. A binned maximum-likelihood analysis of ``FRONT" data~\footnote{\label{front}Again, the reason for adopting only ``FRONT" data is to achieve a compromise between a small PSF and lessened energy dispersion.} in 0.7$-$400 GeV was performed for each individual segment.  The choice of a minimum energy cut of 0.7 GeV in this analysis is motivated by the FWHM of the PSF at 0.7 GeV which is $<$40\% of that at 0.2 GeV (cf. SLAC~$^{\ref{slac}}$). In other words, the source confusion of photons in this analysis is conspicuously \emph{less} serious. We assumed a PL model for \emph{Fermi} J1841.1-0458. We then repeated this procedure with  270 days as a segment. The temporal behaviors of the photon flux and photon index of \emph{Fermi} J1841.1-0458 are plotted with the X-ray outburst history of 1E 1841-045, taken from  \citet{Lin2011}, \citet{Collazzi2015} and references therein~$^{\ref{GCN}}$, altogether in Figure~\ref{lc}. %From MJD55582.655 to MJD56302.655 (in 2011 January 21$-$2013 January 10), there is an apparent flux drop accompanied with a drop in the TS value, and its start seems to synchronize with the start of a series of 10 X-ray outbursts of  1E 1841-045 in 2011 February$-$July \citep[cf.][]{Lin2011}.

If we adopt  180 days as a bin, constant-value functions are sufficient to describe the temporal distributions of flux and spectral index, with $\chi^2<20$ for 14 d.o.f., and \emph{no} bins deviate from the best-fit flat line by $\ge$$2.7\sigma$. If we change the bin size to be  270 days, the temporal distributions of flux and spectral index are also uniform, with $\chi^2<13$ for 9 d.o.f., and there are \emph{no} bins different from the best-fit horizontal line by $\ge$$2.2\sigma$.  More importantly, the errors considered in above fittings are purely statistical. If we take the systematic uncertainties of Galactic diffuse emission model and the effective area \citep[cf.][]{Abdo2013} into consideration, then the significance of variability will further be lowered. Thus, \emph{no} $\gamma$-ray variability of \emph{Fermi} J1841.1-0458 is significantly detected.

\section{Discussion} \label{sec:discuss}

%\subsection{Cosmic-ray origins of \emph{Fermi} J1841.1-0458} \label{J1841}

Our analysis of \emph{Fermi} J1841.1-0458 confirms what \citet{Acero2015b} and \citet{Li2016} discovered with \emph{Fermi} LAT data: the presence of a GeV-bright extended source which is positionally consistent with the Kes 73/1E 1841-045 system. \citet{Li2016} placed a 95\% upper limit of $2.05\times10^{-11}$ erg cm$^{-2}$ s$^{-1}$ on the 0.1$-$10 GeV flux of the magnetar 1E 1841-045, after their attempt to subtract the $\gamma$-ray contribution from SNR Kes 73 which is enclosing the magnetar at its center. This upper limit is approximately  one-fifth of our observed flux for \emph{Fermi} J1841.1-0458. Nevertheless, we take a different approach to discuss possible $\gamma$-ray origins.

\emph{Fermi} J1841.1-0458 radiates strongly in both low-energy (LE; 0.1$-$1 GeV), HE (1$-$10 GeV) and VHE ($>$10 GeV) regimes. Comprehensively considering Figure~\ref{1E_SED} and Table~\ref{spectral1}, we found that  the 0.1$-$10 GeV and 10$-$200 GeV emissions may be respectively dominated by two components of different mechanisms. We will therefore discuss the emission mechanisms of the source using these two energy bands.

We compared the LAT spectral shape of \emph{Fermi} J1841.1-0458 with those of other SNRs younger than or as young as Kes 73, which are tabulated in Table~\ref{youngSNR}. The interpretations based on this comparison will be discussed in the following sub-sections.

\subsection{Relations with   SNR Kes 73} \label{Kes73}

The best-fit 1$-$50 GeV centroid of \emph{Fermi} J1841.1-0458  is almost at the center of Kes 73, while the 95\% error region of this centroid  is mutually exclusive with the two  H II clouds  G27.276+0.148 and G27.491+0.189. This contradicts the MC-interaction scenario where the actual $\gamma$-ray source is the proton collision sites (MCs) rather than the proton acceleration sites (SNR shocks). Whereas, this is within our expectation, as only $\sim$5\% of cosmic rays accelerated in Kes 73 can reach  these two H II clouds, according to the angular sizes of the clouds and their angular separations from  Kes 73 \citep[taken from Figure 1(a) of][]{Tian2008}. Also, Kes 73 is a shell-type SNR which shows \emph{no} signs of being thermal composite, further confirming that the dense medium interacting with Kes 73 is \emph{not} that abundant. Furthermore,  a SNR is generally GeV-brightest at an age within $\sim$3$-$10 kyr \citep[cf.][]{Dermer2013}, while  Kes 73 is only $\lesssim$2.1 kyr old \citep[][]{Kumar2014}.  Therefore, the $\gamma$-ray contribution from Kes 73 should be  limited.

The \emph{Fermi} LAT detection of the GeV counterparts of Cas A and Tycho \citep[][]{Abdo2010e, Zirakashvili2014, Giordano2012, Archambault2017}, respectively $\sim$340 yr old and $\sim$440 yr old SNRs in lack of dense medium in their  neighborhoods \citep[][]{Fesen2006, Tian2011, Kilpatrick2014, Kilpatrick2016}, makes it rather possible for  Kes 73  to be a significant $\gamma$-ray source like Cas A and Tycho. However, if  Kes 73 \citep[at a distance of $\sim$8.5 kpc;][]{Tian2008, Kumar2014} is the only source of energy injection for \emph{Fermi} J1841.1-0458, then its 0.1$-$1 GeV luminosity would be a factor of $>$20 greater than those of  Cas A and Tycho \citep[at  distances of $\sim$3.3 kpc and $<$5 kpc respectively;][]{Alarie2014, Hayato2010, Tian2011}.  Besides, the $\sim$1$-$10 GeV spectrum of \emph{Fermi} J1841.1-0458  is steeper than those of all other young-SNR associated sources in Table~\ref{youngSNR} by  $>$$3\sigma$.
Furthermore, $\lesssim$3000 yr old SNRs  should have $\gamma$-ray luminosities $\lesssim$$10^{35}$ erg s$^{-1}$ \citep[cf.][]{Dermer2013}, while the 0.1$-$10  GeV luminosity of \emph{Fermi} J1841.1-0458 is $\sim$$8.7\times10^{35}$ erg s$^{-1}$, based on the BKPL model fit to the 0.2$-$10 GeV spectrum. If we adopt the simple model where the target material for cosmic-ray interaction at early SNR ages is only the shocked interstellar medium but \emph{not} the SNR shell \citep[cf.][]{Dermer2013}, then it is  \emph{unusual} for Kes 73 to dominate the LE$-$HE (0.1$-$10 GeV)  emission of \emph{Fermi} J1841.1-0458.

%The $\gamma$-ray emission from a SNR is \emph{unlikely} to dramatically change with a short timescale of years.  Therefore, if the apparent drop of the 0.7$-$400  GeV flux in 2011 January 21$-$2013 January 10 is \emph{not} an occasional chance event, then the $>$0.7 GeV emission of \emph{Fermi} J1841.1-0458 is even more \emph{unlikely} to be dominated by Kes 73.

In 10$-$200 GeV, the luminosity of \emph{Fermi} J1841.1-0458 is $\sim$$1.6\times10^{35}$ erg s$^{-1}$, with a hard photon index of  $1.99 \pm 0.22$. Interestingly, the 1$-$300 GeV spectrum of the SNR RX J0852.0-4622 \citep[2400$-$5100 yr old;][]{Allen2015} is well described by a PL model with a similar photon index of $1.85\pm0.06$ \citep[][]{Tanaka2011}, and the 0.5$-$300 GeV spectrum of the SNR RX J1713.7-3946 \citep[$\sim$1600 yr old; cf.][]{Fesen2012, Katsuda2015} is well described by a PL model with an even slightly harder photon index of $1.53\pm0.07$ \citep[][]{Federici2015}. For these two SNRs, any simple one-zone model for particle acceleration does not favor a strong hadronic pion-decay component \citep[][]{Federici2015}, as more-energetic particles are more inclined to lose their energy or to escape from the acceleration sites. Such hard GeV spectra of SNRs can possibly be explained by leptonic cosmic-rays up-scattering soft photons \citep[cf.][]{Tanaka2011, Dermer2013}, and/or more complicated hadronic emission (beyond one-zone models) from a shell of dense gas that is transferred from progenitor stellar winds and is located a short distance upstream of the forward shock \citep[cf.][]{Federici2015}.

Since Kes 73 is as young as RX J1713.7-3946 and is just slightly younger than RX J0852.0-4622,  it is likely that Kes 73  dominates the VHE emission of \emph{Fermi} J1841.1-0458, whose VHE spectrum is as hard as those of RX J0852.0-4622 and RX J1713.7-3946. Both the GeV$-$TeV spectra of RX J0852.0-4622 and RX J1713.7-3946 are well described by  BKPL models with spectral turnovers and peaks at $\sim$300 GeV \citep[][]{Tanaka2011, Federici2015}. HESS observation in the TeV band shows that Kes 73 has a detection significance of $\sim$$6.0\sigma$ at the northern edge of  HESS J1841-055 \citep[cf.][]{Aharonian2008, Bochow2011}. Unfortunately, it is \emph{inappropriate} to examine the spectral connection for \emph{Fermi} J1841.1-0458 and HESS J1841-055, since HESS J1841-055 has an extended feature composed of  Kes 73 as well as several stronger high-energy sources \citep[][]{Aharonian2008, Bochow2011}.  It is unclear whether the VHE spectrum of \emph{Fermi} J1841.1-0458 can be connected to the TeV spectrum of  Kes 73 (after subtracting other components of HESS J1841-055) with a turnover and peak at $\sim$300 GeV, like what is observed in the spectra of RX J0852.0-4622 and RX J1713.7-3946. It is still dubious whether the scenarios explaining the GeV spectra of  RX J0852.0-4622 and RX J1713.7-3946 can be applied to  Kes 73 or not.

\subsection{Relations with the magnetar 1E 1841-045} \label{1E}

1E 1841-045 is highly consistent with the 1$-$50 GeV centroid of \emph{Fermi} J1841.1-0458. One of the most preferable spectral models in 0.2$-$10 GeV, PLE, yields a spectral index within the range of typical values of about 1-2 in the 2nd catalog of pulsars detected by the \emph{Fermi} LAT, and a cutoff energy  consistent with the range of typical values of about 1-4 GeV \citep[][]{Abdo2013}. While the spin-down power of 1E 1841-045 of   $\sim$$1.0\times10^{33}$ erg s$^{-1}$ \citep[][]{Dib2014} is only $\sim$0.11\% of the 0.1$-$10 GeV luminosity of \emph{Fermi} J1841.1-0458, its surface magnetic field of $\sim$$1.4\times10^{15}$ G at the pole \citep[cf.][]{Dib2014} makes it appropriate to estimate the power of magnetic field decay to be $\gtrsim$$10^{36}$ erg s$^{-1}$ \citep[cf. ][]{Zhang2003}, which is marginally greater than the 0.1$-$10 GeV luminosity of \emph{Fermi} J1841.1-0458. Therefore, one possible scenario is that the LE$-$HE emission of \emph{Fermi} J1841.1-0458 is dominated by energy released from the very-intense magnetic field of 1E 1841-045.

The $>$0.7 GeV emission of \emph{Fermi} J1841.1-0458 is essentially steady. Both Figure~\ref{lc} in this work and Figure~5 in \citet{Yeung2016} (a long-term light-curve of the  SGR 1806-20 region) suggest that, regardless of cosmic-ray origins, $\gamma$-ray emissions from magnetars' regions may \emph{not} vary in response to magnetar outbursts, although the enormous energy ejected from these outbursts triggers abrupt yet dramatic enhancements in X-ray and \emph{soft} $\gamma$-ray emissions. %However, we found an apparent `anti-correlation' between LAT flux and magnetar bursts. If the apparent 0.7$-$400  GeV flux drop in 2011 January 21$-$2013 January 10 is \emph{not} a chance event and its start (almost) synchronize with the start of a series of 10 X-ray outbursts of 1E 1841-045 in 2011 February$-$July, then one possible reason is the photon-photon collision between $\gamma$-ray photons and increased X-ray photons following the X-ray outbursts.

The 10$-$200 GeV  luminosity of \emph{Fermi} J1841.1-0458 is 2 orders of magnitude higher than the spin-down power of 1E 1841-045, but is about an order of magnitude lower than its power of magnetic field decay. Although synchrotron cooling generally makes leptonic cosmic-rays difficult to produce $\gamma$-ray photons of a few GeV or above  via synchrotron radiation, PWNe can, with the reduced synchrotron losses for high-energy IC-emitting electrons, maintain their high GeV$-$TeV $\gamma$-ray fluxes for timescales exceeding the lifetime of their progenitor pulsars \citep[][]{Tibolla2011}. Remarkably, in 10$-$200 GeV, the luminosity and spectral shape of \emph{Fermi} J1841.1-0458 are both highly consistent with those of Crab Nebula's IC component \citep[see Table~\ref{youngSNR}, and compare Figure~\ref{1E_SED} in this work with Figure~2 of ][]{Buehler2012}.  Crab Nebula is a PWN powered by a $\sim$1 kyr old pulsar \citep[as young as the Kes 73/1E 1841-045 system; ][]{Rudie2008}, prompting us to speculate that Kes 73 also contains an IC-radiating PWN. On the other hand, a stark contrast between the 0.2$-$10 GeV spectra of \emph{Fermi} J1841.1-0458 and Crab Nebula (cf. Table~\ref{youngSNR}) makes it \emph{unreasonable} to apply the synchrotron component of Crab Nebula \citep[for detail, see][]{Buehler2012} to explain the LE$-$HE emission of \emph{Fermi} J1841.1-0458. 

TeV-detected PWNe should be associated with pulsars releasing power of $>$$10^{36}$ erg s$^{-1}$ \citep[][]{Halpern2010b}. Such required power is greater than the rotational power of 1E 1841-045 by three orders of magnitude, but can be afforded by  magnetic field decay of 1E 1841-045.  Swift J1834.9-0846, the first magnetar confirmed to produce a PWN \citep[][]{Younes2016}, is under a similar circumstance \citep[cf.][]{Kargaltsev2012}, and hence the wind nebula around it has been argued to be magnetically powered \citep[][]{Tong2016}. Conceivably,  magnetar's relativistic outflows, which are accumulated over all outbursts in its history and confined by  SNR shell, can supply even larger fluxes of wind particles than those supplied in quiescence \citep[][]{Granot2017}. Therefore, 1E 1841-045 is sufficient to generate an IC-radiating PWN which may account for a significant portion of the LAT flux in the VHE band and/or the HESS flux in the TeV band. However, a major uncertainty of this scenario is that there is \emph{no} firmly identified PWN inside the SNR shell of Kes 73.

\section{Summary} \label{sec:summary}

Subject to assured authenticity of the model prediction for young ($\lesssim$3 kyr old) SNRs done by \citet{Dermer2013}, the LE$-$HE (0.1$-$10 GeV) emission of \emph{Fermi} J1841.1-0458  \emph{cannot} be dominated by   SNR Kes 73, and is more likely to be dominated by the magnetar 1E 1841-045. Whereas, unless more and more young  SNRs are found to have low  0.1$-$10 GeV luminosities (like those of Cas A, Tycho, RX J0852.0-4622 and RX J1713.7-3946),  the credibility of their model for SNR evolution may still be challenged. Once it is challenged, we \emph{cannot} robustly eliminate the possibility of the Kes 73 origin for this band. The VHE ($>$10 GeV) emission of \emph{Fermi} J1841.1-0458 is evidently dominated by Kes 73, while a  PWN generated by  1E 1841-045, if it exists, is a sufficient yet unnecessary source in this energy band.  %We also found an apparent 0.7$-$400  GeV flux drop in 2011 January 21$-$2013 January 10 (at a $>$97\% confidence level). If this is \emph{not} a chance event, then it can \emph{hardly} be explained in terms of Kes 73, but is possibly related to the X-ray outburst activities of   1E 1841-045.

We suggest modelling the Kes 73/1E 1841-045 system as a single source independent from HESS J1841-055 in the TeV regime, so that we can perform physically meaningful comparisons between GeV and TeV spectra.  In turn, we can compare its GeV$-$TeV spectrum to those of RX J0852.0-4622, RX J1713.7-3946 and Crab Nebula. How will the GeV and TeV analyses of this magnetar/SNR pair be affected if we assign a different morphology to HESS J1841-055? What should be the refined position and dimension of HESS J1841-055 so that it excludes  the magnetar/SNR pair more precisely while retaining all other high-energy components? To address such crucial issues, ones can start with analyzing TeV data collected by HESS, MAGIC and/or CTA, each of which has a higher sensitivity than the GeV-observing \emph{Fermi} LAT.

\section*{Acknowledgement}

This work is supported by the Ministry of Science and Technology of the Republic of China (Taiwan) through grants 103-2628-M-007-003-MY3 and 105-2112-M-007-033-MY2. P.H.T.T. is supported by the One Hundred Talent Program of the Sun Yat-Sen University and the Fundamental Research Funds for the Central Universities in P. R. China. C.Y.H. is supported by the National Research Foundation of Korea through grants 2014R1A1A2058590 and 2016R1A5A1013277. J.T. is supported by NSFC grants of Chinese Government under 11573010 and U1631103. K.S.C. is supported by a GRF grant under 17300814. P.K.H.Y. thanks X. Hou and L. C. C. Lin for some useful discussion.




\clearpage

\clearpage

\clearpage

\clearpage

\begin{table}
	\caption{The $2\Delta$$ln(likelihood)$ in 1$-$400 GeV for ``FRONT" data, when uniform disks of different radii replace the point-source model to be the morphology of \emph{Fermi} J1841.1-0458.}
	\begin{center}
		\begin{tabular}{lc}
			\hline\hline
			Radius of extension (deg) & $2\Delta$$ln(likelihood)$ \\ \hline
			0.1 & 16.2026 \\ 
			0.2 & 42.6688 \\ 
			0.3 & 59.6818 \\ 
			0.31 & 59.8536 \\ 
			0.32 & 60.9114 \\ 
			0.325 & 60.8942 \\ 
			0.33 & 60.8948 \\ 
			0.34 & 60.8702 \\ 
			0.35 & 60.6538 \\ 
			0.36 & 60.7046 \\ 
			0.37 & 59.9428 \\ 
			0.4 & 56.9028 \\ \hline
		\end{tabular}
	\end{center}
	\label{Ext}
\end{table}

\clearpage

\begin{table}
	\caption{$\gamma$-ray spectral properties of \emph{Fermi} J1841.1-0458 as observed  by \emph{Fermi} LAT.}
	\begin{center}
		\begin{tabular}{lcc}
			\hline\hline
			& 0.2$-$10 GeV & 10$-$200 GeV   \\ \hline
			\multicolumn{3}{c}{PL}     \\ \hline
			$\Gamma$ & 2.247 $\pm$ 0.039 & 1.992 $\pm$ 0.223 \\ 
			Flux ($10^{-9}$ photons cm$^{-2}$ s$^{-1}$) & 84.68 $\pm$ 4.28 & 0.362 $\pm$ 0.091 \\ 
			TS & 678.8 & 21.7   \\ \hline
			\multicolumn{3}{c}{PLE}     \\ \hline
			$\Gamma$ & 1.952 $\pm$ 0.132 & 1.832 $\pm$ 0.616 \\ 
			$E_\mathrm{c}$ (MeV) & 5092.0 $\pm$ 2121.2 & 244638.0 $\pm$ 992998.0 \\ 
			Flux ($10^{-9}$ photons cm$^{-2}$ s$^{-1}$) & 80.93 $\pm$ 5.0 & 0.362 $\pm$ 0.091 \\ 
			TS & 691.4 & 21.7   \\ \hline
			\multicolumn{3}{c}{BKPL}     \\ \hline
			$\Gamma_1$ & 1.955 $\pm$ 0.082 & 1.16 $\pm$ 0.551 \\ 
			$\Gamma_2$ & 2.631 $\pm$ 0.109 & 58.896 $\pm$ 410.359 \\ 
			$E_\mathrm{b}$ (MeV) & 1241.3 $\pm$ 248.8 & 40480.9 $\pm$ 28812.3 \\ 
			Flux ($10^{-9}$ photons cm$^{-2}$ s$^{-1}$) & 79.47 $\pm$ 9.52 & 0.337 $\pm$ 0.152 \\ 
			TS & 696.9 & 24.3   \\ \hline
		\end{tabular}
	\end{center}
	\label{spectral1}
\end{table}

\clearpage

\begin{table}[]
	\centering
	\caption{Spectral comparison among young-SNR associated sources detected by \emph{Fermi} LAT.}
	\label{youngSNR}
	\begin{tabular}{lcccc}
		\hline\hline
		Source             & SNR age    & Energy band  & Photon index & References    \\
		& (kyr)      & (GeV)        &              &               \\ \hline
		\emph{Fermi} J1841.1-0458 & 0.75$-$2.1   & 0.2$-$1.24     & 1.95 $\pm$ 0.08  & (1), (2)      \\
		&            & 1.24$-$10      & 2.63 $\pm$ 0.11  &               \\
		&            & 10$-$200       & 1.99 $\pm$ 0.22  &               \\ \hline
		Cas A              & $\sim$0.34 & 0.5$-$50       & 2.0 $\pm$ 0.1    & (3), (4)      \\ \hline
		RX J0852.0-4622    & 2.4$-$5.1    & 1$-$300        & 1.85 $\pm$ 0.06  & (5), (6)      \\ \hline
		RX J1713.7-3946    & $\sim$1.6  & 0.5$-$300      & 1.53 $\pm$ 0.07  & (7), (8), (9) \\ \hline
		Crab Nebula        & $\sim$1.0  & 0.1$-$$\sim$1  & 3.59 $\pm$ 0.07  & (10), (11)    \\
		&            & $\sim$1$-$13.9 & 1.48 $\pm$ 0.07  &               \\
		&            & 13.9$-$300     & 2.19 $\pm$ 0.17  &              \\ \hline
		RCW 103 & $\sim$2.0 & 1$-$300 & 2.0 $\pm$ 0.1 & (12), (13) \\ \hline
		Tycho & $\sim$0.44 & 0.3$-$$\sim$500 & 2.14 $\pm$ 0.09 & (14) \\ \hline
	\end{tabular}

\raggedright
\textbf{References.} (1) this work, (2) \citet{Kumar2014}, (3) \citet{Fesen2006}, (4) \citet{Abdo2010e}, (5) \citet{Allen2015}, (6) \citet{Tanaka2011}, (7) \citet{Fesen2012}, (8) \citet{Katsuda2015}, (9) \citet{Federici2015}, (10) \citet{Rudie2008}, (11) \citet{Buehler2012}, (12) \citet{Carter1997}, (13) \citet{Xing2014}, (14) \citet{Archambault2017}.
\end{table}


\listofchanges


\begin{thebibliography}{}

	\bibitem[Abdo et al.(2009)]{Abdo2009} Abdo, A. A., Ackermann, M., Ajello, M., et al. 2009, ApJL, 706, L1
	\bibitem[Abdo et al.(2010a)]{Abdo2010a} Abdo, A. A., Ackermann, M., Ajello, M., et al. 2010a, ApJ, 718, 348
	\bibitem[Abdo et al.(2010b)]{Abdo2010b} Abdo, A. A., Ackermann, M., Ajello, M., et al. 2010b, Sci, 327, 1103
	\bibitem[Abdo et al.(2010c)]{Abdo2010c} Abdo, A. A., Ackermann, M., Ajello, M., et al. 2010c, ApJ, 712, 459
	\bibitem[Abdo et al.(2010d)]{Abdo2010d} Abdo, A. A., Ackermann, M., Ajello, M., et al. 2010d, ApJL, 725, L73 
	\bibitem[Abdo et al.(2010e)]{Abdo2010e} Abdo, A. A., Ackermann, M., Ajello, M., et al. 2010e, ApJL, 710, L92
	\bibitem[Abdo et al.(2013)]{Abdo2013} Abdo, A. A., Ajello, M., Allafort, A., et al. 2013, ApJS, 208, 17
	\bibitem[Acero et al.(2015)]{Acero2015a} Acero, F., Ackermann, M., Ajello, M. et al. 2015, ApJS, 218, 23
	\bibitem[Acero et al.(2016)]{Acero2015b} Acero, F., Ackermann, M., Ajello, M. et al. 2016, ApJS, 224, 8
	\bibitem[Aharonian et al.(2008)]{Aharonian2008} Aharonian, F., Akhperjanian, A. G., Barres de Almeida, U., et al. 2008, A\&A, 477, 353
	\bibitem[Alarie et al.(2014)]{Alarie2014} Alarie, A., Bilodeau, A., \& Drissen, L. 2014, MNRAS, 441, 2996
	%\bibitem[Allafort et al.(2013)]{Allafort2013} Allafort, A., Baldini, L., Ballet, J., et al. 2013, ApJL, 777, L2
	\bibitem[Allen et al.(2015)]{Allen2015} Allen, G. E., Chow, K., DeLaney, T., et al. 2015, ApJ, 798, 82
	\bibitem[Araya(2015)]{Araya2015} Araya, M. 2015, ApJ, 813, 3
	\bibitem[Archambault et al.(2017)]{Archambault2017} Archambault, S., Archer, A., Benbow, W., et al. 2017, ApJ, 836, 23
	\bibitem[Auchettl et al.(2014)]{Auchettl2014} Auchettl, K., Slane, P., \& Castro, D. 2014, ApJ, 783, 32
	%\bibitem[Bamba et al.(2016)]{Bamba2016} Bamba, A., Sawada, M., Nakano, Y., et al. 2016, PASJ, 68, S5
	\bibitem[Beloborodov \& Thompson(2007)]{Beloborodov2007} Beloborodov, A. M., \& Thompson, C. 2007, ApJ, 657, 967
	\bibitem[Bochow(2011)]{Bochow2011} Bochow, A. 2011, PhD thesis, Ruperto-Carola Univ. Heidelberg
	\bibitem[Buehler et al.(2012)]{Buehler2012} Buehler, R., Scargle, J. D., Blandford, R. D., et al. 2012, ApJ, 749, 26
	\bibitem[Carter et al.(1997)]{Carter1997} Carter, L. M., Dickel, J. R., \& Bomans, D. J. 1997, PASP, 109, 990
	%\bibitem[Case \& Bhattacharya (1998)]{Case1998} Case, G. L., \& Bhattacharya, D. 1998, ApJ, 504, 761
	\bibitem[Castro et al.(2013)]{Castro2013} Castro, D., Slane, P., Carlton, A., \& Figueroa-Feliciano, E. 2013, ApJ, 774, 36
	\bibitem[Cheng \& Zhang(2001)]{Cheng2001} Cheng, K. S., \& Zhang, L. 2001, ApJ, 562, 918
	\bibitem[Collazzi et al.(2015)]{Collazzi2015} Collazzi, A. C., Kouveliotou, C., van der Horst, A. J., et al. 2015, ApJS, 218, 11
	%\bibitem[Cordes \& Lazio(2003)]{Cordes2003} Cordes, J. M. \& Lazio, T. J. W. 2003, astro-ph/0301598
	\bibitem[Dermer \& Powale(2013)]{Dermer2013} Dermer, C. D., \& Powale, G. 2013, A\&A, 553, A34
	\bibitem[Dib \& Kaspi(2014)]{Dib2014} Dib, R., \& Kaspi, V. M. 2014, ApJ, 784, 37
	\bibitem[Duncan(1998)]{Duncan1998} Duncan, R. C. 1998, ApJ, 498, L45
	\bibitem[Duncan \& Thompson(1992)]{Duncan1992} Duncan, R. C., \& Thompson, C. 1992, ApJ, 392, L9
	\bibitem[Federici  et al.(2015)]{Federici2015} Federici, S., Pohl, M., Telezhinsky, I., Wilhelm, A., \& Dwarkadas, V. V. 2015, A\&A, 577, A12
	\bibitem[Fesen et al.(2006)]{Fesen2006} Fesen, R. A., Hammell, M. C., Morse, J., et al. 2006, ApJ, 636, 859
	\bibitem[Fesen et al.(2012)]{Fesen2012} Fesen, R. A., Kremer, R., Patnaude, D., \& Milisavljevic, D. 2012, AJ, 143, 27
	\bibitem[Gaensler \& Slane(2006)]{Gaensler2006} Gaensler, B. M., \& Slane, P. O. 2006, ARA\&A, 44, 17
	%\bibitem[Giacani et al.(2009)]{Giacani2009} Giacani, E., Smith, M. J. S., Dubner, G., et al. 2009, A\&A, 507, 841
	%\bibitem[Ginzburg \& Syrovatskii(1964)]{Ginzburg1964} Ginzburg, V. L., \& Syrovatskii, S. I. 1964, The Origin of Cosmic Rays (New York: Macmillan)
	\bibitem[Giordano et al.(2012)]{Giordano2012} Giordano, F., Naumann-Godo, M., Ballet, J., et al. 2012, ApJ, 744, L2
	\bibitem[Granot et al.(2017)]{Granot2017} Granot, J., Gill, R., Younes, G., et al. 2017, MNRAS, 464, 4895
	\bibitem[Green et al.(1997)]{Green1997} Green, A. J., Frail, D. A., Goss, W. M., \& Otrupcek, R. 1997, AJ, 114, 2058
	%\bibitem[Halpern \& Gotthelf(2010a)]{Halpern2010a} Halpern, J. P., \& Gotthelf, E. V. 2010a, ApJ, 709, 436
	\bibitem[Halpern \& Gotthelf(2010)]{Halpern2010b} Halpern, J. P., \& Gotthelf, E. V. 2010, ApJ, 725, 1384
	\bibitem[Hayato et al.(2010)]{Hayato2010} Hayato, A., Yamaguchi, H., Tamagawa, T., et al. 2010, ApJ, 725, 894
	\bibitem[Helfand et al.(1992)]{Helfand1992} Helfand, D. J., Zoonematkermani, S., Becker, R. H., \& White, R. L. 1992, ApJS, 80, 211
	%\bibitem[Hobbs et al.(2004)]{Hobbs2004} Hobbs, G., Lyne, A. G., Kramer, M., Martin, C. E., \& Jordan, C. A. 2004, MNRAS, 353, 1311 
	%\bibitem[Hui et al.(2016)]{Hui2016} Hui, C. Y., Yeung, P. K. H., Ng, C. W., et al. 2016, MNRAS, 457, 4262
	%\bibitem[Iwamoto et al.(1998)]{Iwamoto1998} Iwamoto, K., Mazzali, P. A., Nomoto, K., et al. 1998, Nature, 395, 672
	%\bibitem[Iwamoto et al.(2000)]{Iwamoto2000} Iwamoto, K., Nakamura, T.; Nomoto, K., et al. 2000, ApJ, 534, 660
	\bibitem[Kargaltsev et al.(2012)]{Kargaltsev2012} Kargaltsev, O., Kouveliotou, C., Pavlov, G. G., et al., 2012, ApJ, 748, 26
	\bibitem[Katsuda et al.(2015)]{Katsuda2015} Katsuda, S., Acero, F., Tominaga, N. et al. 2015, ApJ, 814, 29
	\bibitem[Kilpatrick et al.(2014)]{Kilpatrick2014} Kilpatrick, C. D., Bieging, J. H., \& Rieke, G. H. 2014, ApJ, 796, 144
	\bibitem[Kilpatrick et al.(2016)]{Kilpatrick2016} Kilpatrick, C. D., Bieging, J. H., \& Rieke, G. H. 2016, ApJ, 816,1
	\bibitem[Kumar et al.(2014)]{Kumar2014} Kumar, H. S., Safi-Harb, S., Slane, P. O., \& Gotthelf, E. V. 2014, ApJ, 781, 41
	\bibitem[Li et al.(2017)]{Li2016} Li, J., Rea, N., Torres, D. F., \& de Ona-Wilhelmi, E. 2017, ApJ, 835, 30
	\bibitem[Lin et al.(2011)]{Lin2011} Lin, L., Kouveliotou, C., G\"{o}\u{g}\"{u}\c{s}, E., et al. 2011, ApJL, 740, L16
	\bibitem[Liu et al.(2015)]{Liu2015} Liu, B., Chen, Y., Zhang, X. et al. 2015, ApJ, 809, 102
	\bibitem[Ma et al.(2016)]{Ma2016} Ma, Y. K., Ng, C.-Y., Bucciantini, N., et al. 2016, ApJ, 820, 100
	%\bibitem[Mazzali et al.(2002)]{Mazzali2002} Mazzali, P. A., Deng, J., Maeda, K., et al. 2002, ApJ, 572, L61
	%\bibitem[Mazzali et al.(2006)]{Mazzali2006} Mazzali, P. A., Deng, J., Pian, E., et al. 2006, ApJ, 645, 1323
	%\bibitem[Ng et al.(2016)]{Ng2016} Ng, C. W., Takata, J., \& Cheng, K. S. 2016, ApJ, 825, 18
	\bibitem[Olausen \& Kaspi(2014)]{Olausen2014} Olausen, S. A., \& Kaspi, V. M. 2014, ApJS, 212, 6
	%\bibitem[Rea et al.(2014)]{Rea2014} Rea, N., Vigano', D., Israel, G. L., Pons, J. A., \& Torres, D. F. 2014, ApJL, 781, L17
	\bibitem[Rudie et al.(2008)]{Rudie2008} Rudie, G. C., Fesen, R. A., \& Yamada, T. 2008, MNRAS, 384, 1200
	%\bibitem[Scoville et al.(1987)]{Scoville1987} Scoville, N. Z., Yun, M. S., Sanders, D. B., Clemens, D. P., \& Waller, W. H. 1987, ApJS, 63, 821
	%\bibitem[Seward et al.(2003)]{Seward2003} Seward, F. D., Slane, P. O., Smith, R. K., et al. 2003, ApJ, 584, 414
	\bibitem[Slysh et al.(1999)]{Slysh1999} Slysh, V. I., Val'tts, I. E., Kalenskii, S. V., et al. 1999, A\&AS, 134, 115
	%\bibitem[Stanimirovi\'c et al.(2003)]{Stanimirovic2003} Stanimirovi\'c, S., Weisberg, J. M., Dickey, J. M., et al. 2003, ApJ, 592, 953
	%\bibitem[Sun et al.(2004)]{Sun2004} Sun, M., Seward, F. D., Smith, R. K., \& Slane, P. O. 2004, ApJ, 605, 742
	\bibitem[Takata et al.(2013)]{Takata2013} Takata, J., Wang, Y., Wu, E. M. H., Cheng, K. S., 2013, MNRAS, 431, 2645
	\bibitem[Tanaka et al.(2011)]{Tanaka2011} Tanaka, T., Allafort, A., Ballet, J., et al. 2011, ApJL, 740, L51
	\bibitem[Tian \& Leahy(2008)]{Tian2008} Tian, W. W., \& Leahy, D. A. 2008, ApJ, 677, 292
	\bibitem[Tian \& Leahy(2011)]{Tian2011} Tian, W. W., \& Leahy, D. A. 2011, ApJL, 729, L15
	\bibitem[Tibolla et al.(2011)]{Tibolla2011} Tibolla, O., Mannheim, K., Elsässer, D., \& Kaufmann, S. 2011, arXiv:1111.1634
	\bibitem[Tong(2016)]{Tong2016} Tong, H. 2016, RAA, 16, 143
	\bibitem[Vasisht \& Gotthelf(1997)]{Vasisht1997} Vasisht, G., \& Gotthelf, E. V. 1997, ApJL, 486, L129
	\bibitem[Wachter et al.(2004)]{Wachter2004} Wachter, S., Patel, S. K., Kouveliotou, C., et al. 2004, ApJ, 615, 887
	%\bibitem[Wu et al.(2017)]{Wu2017} Wu, C. H., Ng, C. W., Hou, X., et al. 2017, ApJ, in preparation
	\bibitem[Xing et al.(2014)]{Xing2014} Xing, Y., Wang, Z., Zhang, X., \& Chen, Y. 2014, ApJ, 781, 64
	\bibitem[Yeung et al.(2016)]{Yeung2016} Yeung, P. K. H., Kong, A. K. H., Tam, P. H. T., et al. 2016, ApJ, 827, 41
	\bibitem[Younes et al.(2016)]{Younes2016} Younes, G., Kouveliotou, C., Kargaltsev, O., et al. 2016, ApJ, 824, 138
	\bibitem[Zhang(2003)]{Zhang2003} Zhang, B. 2003, in Proc. Int. Workshop on Strong Magnetic Fields and Neutron Stars: Spin-Down Power of Magnetars, ed. H. J. Mosquera Cuesta, H. Per\'ez Rojas, \& C. A. Zen Vasconcellos, 83
	%\bibitem[Zhou et al.(2014)]{Zhou2014} Zhou, P., Chen, Y., Li, X.-D., et al. 2014, ApJL, 781, L16
	\bibitem[Zirakashvili et al.(2014)]{Zirakashvili2014} Zirakashvili, V. N., Aharonian, F., Yang, R., Wilhelmi, E. O., \& Tuffs, R. J. 2014, ApJ, 785, 130
	%\bibitem[Zubrin \& Shulga(2008)]{Zubrin2008} Zubrin, S. Y., \& Shulga, V. M. 2008, in Young Scientists 15th Proceedings, ed. V. Y. Choliy \& G. Ivashchenko, 41$-$43

\end{thebibliography}
\end{document}